\documentclass[aip,graphicx,eqsecnum]{revtex4-2}

\usepackage{graphicx}
\usepackage{dcolumn}
\usepackage{bm}

\renewcommand{\baselinestretch}{1}

\newcommand{\beq}{\begin{equation}}
\newcommand{\eeq}{\end{equation}}
\newcommand{\beqa}{\begin{eqnarray}}
\newcommand{\eeqa}{\end{eqnarray}}
\newcommand{\vc}[1]{\mbox{\boldmath $#1$}}

\newcommand{\vol}[1]{{\bf #1}}

\newcommand{\du}[1]{{\bf\sf #1}}

\begin{document}


\title{Optimizing the Mean Swimming Velocity of a Model Two-sphere Swimmer}

\author{B. U. Felderhof}

 \email{ufelder@physik.rwth-aachen.de}
\affiliation{Institut f\"ur Theorie der Statistischen Physik\\ RWTH Aachen University\\
Templergraben 55\\52056 Aachen\\ Germany\\
}%

\date{\today}

\begin{abstract}
The swimming of a two-sphere system oscillating in a viscous fluid
is studied on the basis of simplified equations of motion which
take account of both friction and inertial effects. In the model
the friction follows from an Oseen approximation to the mobility
matrix, and the inertial effects follow from a dipole
approximation to the added mass matrix. The resulting mean
swimming velocity is evaluated analytically in a first harmonics
approximation. For specific choices of the parameters this is
compared with the exact result following from a numerical
calculation including higher harmonics. The Oseen-Dipole model is compared with the simpler
Oseen* model, in which the added mass effects are approximated by
just the effective mass of the single spheres and dipole interactions are neglected. The expression for the mean swimming velocity can be reduced to a dimensionless scaling form. For given viscosity and mass density of the fluid the frequency of the stroke and the ratio of radii can be chosen such that the
swimming velocity is optimized.
\end{abstract}

\maketitle
\section{\label{I}Introduction}

It was shown some years ago in experiment and computer simulation by Klotsa et al. \cite{1} that
a system of two neutrally buoyant spheres, oscillating relative to each other along the
axis connecting the centers, will swim when immersed in
a viscous fluid, provided that the two radii differ. Subsequently we showed that for a
class of linear chain models with chosen hydrodynamic interactions, the mean swimming velocity
can be evaluated from an exact expression \cite{2}. Unfortunately, in the expression the
dependence on parameters is not explicit, and this makes it hard to get qualitative insight.
In the following we consider two two-sphere models in more detail, the Oseen$^*$ model and the
Oseen-Dipole model.

In the Oseen model frictional hydrodynamic interactions are treated in Oseen approximation,
and added mass effects are neglected. In the Oseen$^*$ model the mass of the spheres is replaced by the effective mass of the single spheres.
In the Oseen-Dipole model, in addition to the Oseen interaction,
added mass effects are treated in dipole approximation. We shall show that a first harmonics
approximation is accurate and provides qualitative information.

Due to the scallop theorem \cite{3} the mean
swimming velocity vanishes in the Stokes limit of high viscosity in all three models. The behavior
beyond this limit in the Oseen model was studied recently by Hubert et al. \cite{4} to
second order in the amplitude of the stroke.
We commented that the mean swimming velocity can be evaluated in a first harmonics approximation in
close agreement with the exact result, and that when expanded to second order in the amplitude and for large center-to-center distance,
this agrees with the second order expression \cite{5}.
Hubert et al. \cite{4} also performed experiments and lattice Boltzmann simulations, and
showed good agreement with their small amplitude theory.

In the following we investigate the effect of inertia on the swimming of a two-sphere on
the basis of the Oseen$^*$ model and the Oseen-Dipole model.
We show
that the analytic expression for the mean swimming
velocity derived in first harmonics approximation is practically useful for qualitative
exploration.In first harmonics approximation only zeroth and first harmonics are taken into account. The results can be compared with the exact ones, obtained by taking account of higher harmonics in a numerical calculation.

We consider two spheres of radii $a$ and $b$ immersed in a viscous
incompressible fluid of shear viscosity $\eta$ and mass density
$\rho$. The spheres are assumed to be uniform with mass densities
$\rho_a$ and $\rho_b$. We consider situations where the sphere
centers perform harmonic oscillations along the center-to-center line, which is taken to be
the $x$ axis of a Cartesian system of coordinates. It suffices to
consider the coordinates $x_1(t)$ and $x_2(t)$ of the two centers.
The model calculations are based on postulated equations of motion
\cite{6} for the two center positions $\du{R}=(x_1,x_2)$ and
velocities $\du{U}=(U_1,U_2)$. The equations involve interaction
forces, actuating forces, and kinetic forces deriving from the
position dependence of the kinetic energy of flow \cite{7}, as
expressed in the mass matrix. We take a kinematic point of view
and assume that the relative position $x(t)=x_2(t)-x_1(t)$ is
prescribed as
\begin{equation}
\label{1.1}x(t)=d+f\sin(\omega t).
\end{equation}
This suffices to determine the corresponding periodic solution of the equations of motion.
If desired, the various forces can eventually be calculated from the solution.

We are interested in the asymptotic periodic motion of the center $C(t)=(x_1(t)+x_2(t))/2$.
The average of the velocity $U(t)=dC/dt$ over a period at long times defines the mean
swimming velocity $\overline{U_{sw}}$. We write this as
\begin{equation}
\label{1.2}\overline{U_{sw}}=Va\omega,
\end{equation}
with a dimensionless factor $V$, which is called the velocity function. We note that $2\pi V$ is
the fraction of $a$ traversed in
a period of time $T=2\pi/\omega$.
We show that the first harmonics approximation $V^{(1)}$
is a function of the length ratios $\xi=b/a,\;\delta=d/a,\;\varepsilon=f/a$, and the mass density ratios
$\rho_b/\rho_a$ and $\rho/\rho_a$.
It depends in addition on stroke frequency
$\omega$, viscosity $\eta$,
and mass density $\rho$ via the dimensionless combination $R=a^2\omega\rho/\eta$.
We focus in particular on the question of how to optimize $V^{(1)}$ by variation of
parameters.

We expect`that the Oseen-Dipole model catches the main
features of the physical two-sphere swimmer. It provides an approximate account of both friction
and added mass effects. However, the
numerical solution of the Navier-Stokes equations by Dombrowski
and Klotsa \cite{8} showed that at high frequency the system shows
a transition with a reversal of the direction of swimming. Recently Derr et al. \cite{9} analyzed the flow about an oscillating dimer at
small amplitudes on the basis of the Navier-Stokes equations. They found a reversal of swimming velocity at high frequency due to a boundary layer effect, similar to that found by Riley for steady streaming about a single sphere \cite{10}. Hubert et
al. \cite{4} did not see the transition in their lattice Boltzmann
simulations. It would clearly be of interest to modify the
hydrodynamic interactions of our mechanical model in such a way as
to reproduce the reversal of velocity. Such
an investigation is beyond the scope of the present work.

\section{\label{II}Oseen-Dipole two-sphere swimmer}

The so-called Oseen-Dipole two-sphere swimmer is a simple model of a physical two-sphere swimmer consisting of two spheres immersed in an incompressible viscous fluid oscillating relative to each other along a common axis. In the model the hydrodynamic
interactions between the two spheres are approximated by instantaneous point interactions calculated from  an Oseen approximation to the $2\times 2$ mobility matrix
$\vc{\mu}$ and a dipole approximation to the $2\times 2$ mass matrix $\du{m}$. Higher order multipoles and forces of the Basset type are neglected.

The inverse of the Oseen mobility matrix yields a corresponding approximation to the friction matrix
$\vc{\zeta}$. Explicitly the friction matrix is given by
\begin{equation}
\label{2.1}\vc{\zeta}=\frac{12\pi\eta x}{4x^2-9ab}\left(\begin{array}{cc}2ax&-3ab\\-3ab&
2bx\end{array}\right).
\end{equation}
Here the time-dependent distance $x$ is given by Eq. (1.1).

The approximation to the mass matrix reads
\begin{equation}
\label{2.2}\du{m}=\frac{2\pi}{3x^6-3a^3b^3}\left(\begin{array}{cc}a^3[x^6(\rho+2\rho_a)
+2a^3b^3(\rho-\rho_a)]&-3a^3b^3x^3\rho\\-3a^3b^3x^3\rho&
b^3[x^6(\rho+2\rho_b)+2a^3b^3(\rho-\rho_b)]\end{array}\right).
\end{equation}
This also depends on time via Eq. (1.1). At large distance the matrix tends to the time-independent form
\begin{equation}
\label{2.3}\du{m}_\infty=\left(\begin{array}{cc}m_1^*&0\\0&m_2^*\end{array}\right)
\end{equation}
with effective masses
\begin{equation}
\label{2.4}m_1^*=\frac{2\pi}{3}\;a^3(2\rho_a+\rho),\qquad m_2^*=\frac{2\pi}{3}\;b^3(2\rho_b+\rho).
\end{equation}
Here the second term in each case is called the added mass. This represents the kinetic energy of dipolar flow about a moving sphere \cite{7}.

The center velocity $U(t)=dC/dt$ satisfies the equation of motion \cite{6}
\begin{equation}
\label{2.5}\frac{d}{dt}\;(MU)+ZU=\mathcal{I}
\end{equation}
with time-dependent mass $M(t)$ and friction coefficient $Z(t)$ given by
 \begin{equation}
\label{2.6}M=\du{u}\cdot\du{m}\cdot\du{u},\qquad Z=\du{u}\cdot\vc{\zeta}\cdot\du{u},\qquad\du{u}=(1,1),
\end{equation}
and impetus $\mathcal{I}(t)$ given by
\begin{equation}
\label{2.7}\mathcal{I}(t)=-\frac{d}{dt}(\du{u}\cdot\du{m}\cdot\dot{\du{d}})
-\du{u}\cdot\vc{\zeta}\cdot\dot{\du{d}},
\end{equation}
where $\dot{\du{d}}$ is the time-derivative of the
displacement vector $\du{d}(t)$,
\begin{equation}
\label{2.8}\du{d}=\frac{1}{2}(-x+d,x-d).
\end{equation}

The time-dependent mass is found to be
\begin{equation}
\label{2.9}M(t)=m_1^*+m_2^*-2\pi\rho a^3b^3\frac{2x^3-a^3-b^3}{x^6-a^3b^3},
\end{equation}
with $x(t)$ given by Eq. (1.1).
The friction coefficient is found as
\begin{equation}
\label{2.10}Z(t)=24\pi\eta x\frac{(a+b)x-3ab}{4x^2-9ab}.
\end{equation}

The impetus $\mathcal{I}(t)$ is a sum of two terms,
\begin{equation}
\label{2.10}\mathcal{I}(t)=\mathcal{I}_m(t)+\mathcal{I}_f(t),
\end{equation}
with mass contribution $\mathcal{I}_m(t)$ and friction contribution $\mathcal{I}_f(t)$. The mass
contribution is written conveniently in the form
\begin{equation}
\label{2.11}\mathcal{I}_m(t)=\frac{dG_m}{dt},
\end{equation}
with $G_m(t)$ given by
\begin{equation}
\label{2.12}G_m(t)=\frac{1}{2}f\omega(m_1^*-m_2^*)\cos(\omega t)
-\pi f\rho\omega\frac{a^3b^3(a^3-b^3)}{a^3b^3-x^6}\cos(\omega t).
\end{equation}
The friction contribution is
\begin{equation}
\label{2.13}\mathcal{I}_f(t)=12\pi\eta\omega(a-b)f\cos(\omega t)\frac{x^2}{4x^2-9ab}.
\end{equation}
We note that this can be expressed as
\begin{equation}
\label{2.14}\mathcal{I}_f(t)=\frac{dG_f}{dt}
\end{equation}
with periodic function $G_f(t)$. Explicitly
\begin{equation}
\label{2.15}G_f(t)=3\pi\eta(a-b)\bigg[x+\frac{3}{4}\sqrt{ab}
\log\frac{(3\sqrt{ab}+2d)(3\sqrt{ab}-2x)}{(3\sqrt{ab}-2d)(3\sqrt{ab}+2x)}\bigg].
\end{equation}
It follows from Eqs. (2.12) and (2.15) that the time average of $\mathcal{I}(t)$ over a period vanishes. It follows from this property and from Eq. (2.5)
that for the periodic solution $U(t)$ the time-average of the drag force $Z(t)U(t)$
over a period vanishes: $\overline{ZU}=0$, in agreement with general theory \cite{6}. In the following we discuss the solution of Eq. (2.5) and the corresponding calculation of the mean swimming velocity
$\overline{U_{sw}}$.

\section{\label{III}First harmonics approximation}

In earlier work \cite{2} we derived an integral expression for the swim velocity $U_{sw}(t)$
corresponding to the asymptotic periodic solution of the swim equation Eq. (2.5). Hence we
derived an expression for the mean swimming velocity $\overline{U_{sw}}$, the average of $U_{sw}(t)$
 over a period $T=2\pi/\omega$.

The expressions allow exact calculation, but do not provide easy information on the dependence
on parameters. Here we describe an alternative method which provides more insight and leads to
a faster numerical scheme. In the new scheme we solve Eq. (2.5) by expansion in harmonics. The
periodic swim velocity $U_{sw}(t)$ is expanded as
\begin{equation}
\label{3.1}U_{sw}(t)=U_0+\sum^\infty_{n=1}\big[U_{nc}\cos(n\omega t)+U_{ns}\sin(n\omega t)\big].
\end{equation}
Clearly $\overline{U_{sw}}=U_0$. In $n$th harmonics approximation we include harmonics of orders 0 to $n$, and ignore higher
orders. The calculation converges rapidly with increasing $n$. Moreover, it turns out that already
the first harmonics approximation with $n=1$ is numerically close to the exact result.

We examine the first harmonics approximation in analytic form. In this approximation we have
\begin{equation}
\label{3.2}U_{sw}^{(1)}(t)=U_0^{(1)}+U_{1c}^{(1)}\cos(\omega t)+U_{1s}^{(1)}\sin(\omega t).
\end{equation}
Substituting into Eq. (2.5) and expanding the time-dependent factors as in Eq. (3.2)
we obtain a set of three linear equations for the three coefficients
$U_0^{(1)},\;U_{1c}^{(1)},\;U_{1s}^{(1)}$ reading
\begin{eqnarray}
\label{3.3}2Z_0U_0^{(1)}+Z_{1c}U_{1c}^{(1)}+Z_{1s}U_{1s}^{(1)}&=&2\mathcal{I}_0,\nonumber\\
(Z_{1c}+\omega M_{1s})U_0^{(1)}+Z_0U_{1c}^{(1)}+\omega M_0U_{1s}^{(1)}&=&\mathcal{I}_{1c},\nonumber\\
(Z_{1s}-\omega M_{1c}U_0^{(1)}-\omega M_0U_{1c}^{(1)}+Z_0U_{1s}^{(1)}&=&\mathcal{I}_{1s},
\end{eqnarray}
with coefficients which can be calculated from the above.
The equations simplify by use of the properties
\begin{equation}
\label{3.4}\mathcal{I}_0=0,\qquad Z_{1c}=0,\qquad M_{1c}=0.
\end{equation}
From the solution of Eq. (3.3) we find in particular
\begin{equation}
\label{3.5}\overline{U_{sw}}^{(1)}=\frac{-Z_{1s}(Z_0\mathcal{I}_{1s}+\omega
M_0\mathcal{I}_{1c})}{2Z_0^3-Z_0Z_{1s}^2+2\omega^2M_0^2Z_0-\omega^2M_0M_{1s}Z_{1s}}.
\end{equation}
We also find the solution for $U_{1c}^{(1)}$ and $U_{1s}^{(1)}$.

For the mean friction coefficient $Z_0$ we find
\begin{equation}
\label{3.6}Z_0=3\pi\eta\bigg[2(a+b)-3\alpha\beta\frac{(\alpha+\beta)^2}{q_+}
+3\alpha\beta\frac{(\alpha-\beta)^2}{q_-}\bigg],
\end{equation}
with the abbreviations $\alpha=\sqrt{a},\;\beta=\sqrt{b}$, and
\begin{equation}
\label{3.7}q_\pm=\sqrt{4d^2-4f^2+9ab\pm12d\sqrt{ab}}.
\end{equation}
For the coefficient $Z_{1s}$ we find
\begin{equation}
\label{3.8}Z_{1s}=\frac{9\pi\eta\alpha\beta}{f}\bigg[\frac{(\alpha
+\beta)^2}{q_+}(2d+3\alpha\beta)
-\frac{(\alpha-\beta)^2}{q_-}(2d-3\alpha\beta)-4\alpha\beta\bigg].
\end{equation}

The coefficients $\mathcal{I}_{1c}$ and $\mathcal{I}_{1s}$ can be expressed
as a sum of mass and friction contributions
\begin{equation}
\label{3.9}\mathcal{I}_{1c}=\mathcal{I}_{1cm}+\mathcal{I}_{1cf},
\qquad\mathcal{I}_{1s}=\mathcal{I}_{1sm}+\mathcal{I}_{1sf},
\end{equation}
corresponding to the two terms in Eq. (2.7). The coefficients
$\mathcal{I}_{1cm}$ and $\mathcal{I}_{1sf}$ vanish. The coefficient $\mathcal{I}_{1c}$ is given by
\begin{equation}
\label{3.10}\mathcal{I}_{1c}=\frac{3\pi\eta\omega}{4f}(a-b)\big[4f^2-18ab
+3\sqrt{ab}(q_+-q_-)\big].
\end{equation}

The coefficients
$\mathcal{I}_{1sm}$, $M_0$ and $M_{1s}$ must be calculated from integrals
over a period after substitution of $x(t)$ from Eq. (1.1).
We evaluate the required integrals by regarding $x$ as a complex variable and
decomposing the expression into partial fractions. With the definition
\begin{equation}
\label{3.11}w_j=\exp(2j\pi i/6),\qquad j=(1,...,6)
\end{equation}
we have
\begin{equation}
\label{3.12}\frac{1}{g^6-x^6}=\frac{1}{6g^5}\sum^6_{j=1}\frac{w_j}{w_jg-x}.
\end{equation}
Correspondingly we find in particular for the coefficient $\mathcal{I}_{1sm}$
\begin{equation}
\label{3.13}\mathcal{I}_{1sm}=-\frac{1}{2}M_d\omega^2f+J_{1sm},
\end{equation}
with $M_d=m_1^*-m_2^*$ and last term
\begin{equation}
\label{3.14}
J_{1sm}=\frac{\pi}{3}\rho\omega^2\;\alpha\beta(\alpha^6-\beta^6)
\sum^6_{j=1}w_jQ(f,d+w_j\alpha\beta),
\end{equation}
with the function
\begin{equation}
\label{3.15}Q(f,x)=\frac{1}{f}\big[x-\sqrt{x^2-f^2}\big].
\end{equation}
We note that $Q(f,x)\approx f/(2x)$ for $f<<x$, and that it is odd in $f$.

By the same method we find for the mean total mass
\begin{equation}
\label{3.16}M_0=m_1^*+m_2^*+M_{0int},
\end{equation}
where $M_{0int}$ is due to hydrodynamic interactions between the spheres. We find
\begin{equation}
\label{3.17}M_{0int}=\frac{-\pi}{3}\;\rho\alpha\beta\sum^6_{j=1}\big(\alpha^3+(-1)^j\beta^3\big)^2\;\frac{w_j}{W_j},
\end{equation}
with
\begin{equation}
\label{3.18}W_j=\big[d^2-f^2+2w_jd\alpha\beta+w_j^2\alpha^2\beta^2\big]^{1/2}.
\end{equation}
Similarly
\begin{eqnarray}
\label{3.19}M_{1s}=\frac{2\pi}{3f}\rho\alpha\beta\sum^6_{j=1}\big(\alpha^3+(-1)^j\beta^3\big)^2\;\frac{w_j}{W_j}
\big(d+w_j\alpha\beta\big).
\end{eqnarray}
We note that the coefficients $M_{0int},M_{1s}$ are independent of the mass densities $\rho_a,\rho_b$ of the spheres,
and the dependence of $\mathcal{I}_{1s}$ is simple. This can be used to advantage in the optimization of the mean
swimming velocity.

\section{\label{IV}Scaling in first harmonics approximation}

The added mass effect significantly modifies the swimming performance for given stroke. In order to get some
qualitative understanding it is worthwhile to consider the first harmonics approximation. In Eq. (3.5)
for the mean swimming velocity we put
\begin{eqnarray}
\label{4.1} Z_0&=&\eta az_0,\qquad Z_{1s}=\eta a z_{1s},\qquad M_0=\rho a ^3m_0,\nonumber\\
M_d&=&\rho a^3m_d,\qquad M_{1s}=\rho a^3 m_{1s},\qquad M_{0int}=\rho a^3m_{0i},\nonumber\\
\mathcal{I}_{1cf}&=&\eta\omega a^2j_{1cf},\qquad
\mathcal{I}_{1sm}=-\frac{1}{2}f(m_1^*-m_2^*)\omega^2+R\eta\omega a^2j_{1sm},
\end{eqnarray}
where $R=a^2\omega\rho/\eta$. By comparison with Eq. (3.13)
\begin{equation}
\label{4.2}j_{1sm}=\frac{\pi}{3}
\frac{\alpha\beta(\alpha^6-\beta^6)}{a^4}\sum^6_{j=1}w_jQ(f,d+w_j\alpha\beta).
\end{equation}
 The lower case coefficients $z_0,\;z_{1s},\;j_{1cf},\;j_{1sm},\;m_{1s}$
are complicated dimensionless functions of the length ratios $b/a,\;d/a,\;f/a$. The mass
coefficient $m_0$ involves the single sphere bare and added masses, as well as
the added mass corresponding to $M_{0int}$. The dimensionless velocity function,
defined by $V=\overline{U_{sw}}/(a\omega)$, is given in first harmonics approximation by
\begin{equation}
\label{4.3}V^{(1)}=\frac{1}{3}z_{1s}\;\frac{
\pi\varepsilon z_0 (1-\xi^3+2\sigma_a-2\xi^3\sigma_b)-3z_0j_{1sm}-3m_0j_{1cf}}
{z_0(2z_0^2-z_{1s}^2)+m_0(2m_0z_0-m_{1s}z_{1s})R^2}\;R,
\end{equation}
with dimensionless coefficients
 \begin{equation}
\label{4.4}\varepsilon=f/a,\qquad\xi=b/a,\qquad\sigma_a=\rho_a/\rho=1/S_a,\qquad\sigma_b=\rho_b/\rho.
\end{equation}
The velocity function $V^{(1)}$ depends in simple fashion on viscosity, frequency and fluid mass density via the dimensionless scaling variable
$R=a^2\omega\rho/\eta$. The dependence on $\sigma_a,\sigma_b$ is also simple.
The dependence on the four lengths $a,b,d,f$ is complicated. The expression Eq. (4.3) shows that
the mean swimming velocity is generated by an intricate interplay of the effects of friction,
mass, and impetus.

We write Eq. (4.3) in the abbreviated form
\begin{equation}
\label{4.5}V^{(1)}=\frac{AR}{B+CR^2}.
\end{equation}
As a function of the number $R$ this takes its maximum value $V_x^{(1)}$ at
\begin{equation}
\label{4.6}R^{(1)}_x=\sqrt{\frac{B}{C}},
\qquad V^{(1)}_x=\frac{A}{2B}R^{(1)}_x.
\end{equation}
The half-width
of the curve $V^{(1)}(R)$ is
\begin{equation}
\label{4.7}\Delta V^{(1)}=2\sqrt{3}R^{(1)}_{x}.
\end{equation}
We can rewrite Eq. (4.5) as
\begin{equation}
\label{4.8}
\qquad V^{(1)}=2\frac{R^{(1)}_xV^{(1)}_xR}{{R^{(1)}_x}^2+R^2}.
\end{equation}
The two ratios $A/B$ and $A/C$ correspond to the behavior of the mean swimming velocity $\overline{U_{sw}}$
 of a two-sphere as a function of frequency $\omega$ in a fluid of kinematic viscosity $\eta/\rho$
 according to
\begin{eqnarray}
\label{4.9}\overline{U_{sw}}&\approx & 2\frac{a^3\omega^2\rho}{\eta}\;\frac{V^{(1)}_x}{R^{(1)}_x}\;
\;\mathrm{as}\;\;\omega\rightarrow 0,\nonumber\\
\overline{U_{sw}}&\approx & 2 \frac{\eta}{a\rho}R^{(1)}_xV^{(1)}_x\;\;\mathrm{as}\;\;\omega\rightarrow\infty.
\end{eqnarray}

It is interesting to observe that the coefficient $B$
depends only on friction coefficients, that the coefficient $C$ arises from a combination
of mass and friction, and that the coefficient $A$ involves a combination of friction,
mass, and impetus. The first term in the numerator of the expression in Eq. (4.3) is a
measure of the asymmetry of the two-sphere. In addition there are the two terms with moments
of the impetus. The precise combination of terms in Eq. (4.3) follows from the details
of the equations of motion. It would be difficult to write down the expression on intuitive
grounds.

We comment on the simple frequency-dependence found in Eq. (4.3). Consider the linear
susceptibility $\chi(\omega)$ corresponding to a single exponential relaxation process,
\begin{equation}
\label{4.9}\chi(\omega)=\chi_0\int^\infty_0e^{i\omega t}\gamma e^{-\gamma t}dt.
\end{equation}
This is given by
\begin{equation}
\label{4.10}\chi(\omega)=\chi_0\frac{\gamma}{\gamma-i\omega}.
\end{equation}
The absorption is given by the imaginary part
\begin{equation}
\label{4.11}\chi''(\omega)=\chi_0\frac{\gamma\omega}{\gamma^2+\omega^2}.
\end{equation}
This is precisely the frequency-dependence seen in Eq. (4.5). The mathematical
equivalence does not allow us to conclude that there is a relaxation process involved in the
mechanism of swimming. However, it does suggest how to modify the hydromechanical model to
get agreement with the behavior seen in the numerical calculations by
Dombrowski and Klotsa \cite{8}.

We can generalize Eq. (4.11) by adding a second pole on the imaginary $\omega$-axis to get
the susceptibility
\begin{equation}
\label{4.12}\chi_2(\omega)=\chi_0\frac{\gamma}{\gamma-i\omega}
+\chi_{02}\frac{\gamma_2}{\gamma_2-i\omega}.
\end{equation}
In Fig. 1 we plot $\chi''_2(\omega)$ and compare with $\chi''(\omega)$ as functions of $\omega$
for the numerical example $\chi_0=1,\;\gamma=1,\;\chi_{02}=-1,\;\gamma_2=10$.  This shows that
the addition of the second pole with negative residue leads to the type of behavior seen by
Dombrowski and Klotsa \cite{8}. The hydrodynamic interactions in our hydromechanical model
must be modified to reproduce this behavior in the high frequency regime.

\section{\label{V}Expansion in amplitude}

The dimensionless amplitude of the stroke is defined as the ratio
$\varepsilon=f/a$. Since the mean swimming velocity cannot depend
on the phase of the stroke the velocity function $V$ must be an even function of
$\varepsilon$. The first few terms of the Taylor expansion read
\begin{equation}
\label{5.1}V=v_2\varepsilon^2+v_4\varepsilon^4+O(\varepsilon^6).
\end{equation}
However, in first harmonic approximation it is clearly advantageous to keep the
analytic structure in the variable $R$,
as given by Eq. (4.5). The number $R=a^2\omega\rho/\eta$ contains the properties of the fluid,
whereas the other variables refer to the
spheres. In the following we derive expressions for the first two terms in the
expansion of the various coefficients
in Eq. (4.3) in powers of the amplitude $\varepsilon$.

First we consider the friction coefficient. From Eq. (3.6) we find
\begin{equation}
\label{5.2}z_0=z_{0,0}+\varepsilon^2 z_{0,2}+O(\varepsilon^4),
\end{equation}
with first term
\begin{equation}
\label{5.3}z_{0,0}=6\pi\frac{a+b}{a}-18\pi\frac{b(4d-3a-3b)}{4d^2-9ab},
\end{equation}
and second term
\begin{equation}
\label{5.4}z_{0,2}=36\pi\frac{a^2b}{(4d^2-9ab)^3}\big[(36d^2+27ab)(a+b)-108abd-16d^3\big].
\end{equation}
For the friction moment $z_{1s}$ we find
\begin{equation}
\label{5.5}z_{1s}=\varepsilon z_{1s,1}+\varepsilon^3 z_{1s,3}+O(\varepsilon^5),
\end{equation}
with coefficients
\begin{eqnarray}
\label{5.6}z_{1s,1}&=&72\pi ab\frac{(2 d-3a)(2 d-3b)}{(4d^2-9ab)^2},\nonumber\\
z_{1s,3}&=&216\pi\frac{a^3b}{(4d^2-9ab)^4}\big[81a^2b^2-108abd(a+b)+216abd^2
-48(a+b)d^3+16d^4\big].\nonumber\\
\end{eqnarray}

Next we consider the mass. For the added mass in Eq. (3.17) we find
\begin{equation}
\label{5.7}m_{0int}=m_{0i,0}+\varepsilon^2 m_{0i,2}+O(\varepsilon^4),
\end{equation}
with the coefficient
\begin{equation}
\label{5.8}m_{0i,0}=-2\pi\frac{b^3(a^3+b^3-2d^3)}{a^3b^3-d^6},
\end{equation}
and with
\begin{equation}
\label{5.9}m_{0i,2}=3\pi\frac{a^2
b^3d}{(a^3b^3-d^6)^3}\big[2a^6b^6-5a^3b^3(a^3+b^3)d^3+18a^3b^3d^6-7(a^3+b^3)d^9+4d^{12}\big].
\end{equation}
For the total mass we have
\begin{equation}
\label{5.10}M_{0,0}=m_1+\frac{1}{2}m_{1f}+m_2+\frac{1}{2}m_{2f}+\rho a^3m_{0i,0},
\qquad m_{0,2}=m_{0i,2}.
\end{equation}
For the moment $M_{1s}=\rho a^3m_{1s}$ in Eq. (3.19) we find the expansion
\begin{equation}
\label{5.11}m_{1s}=m_{1s,1}\varepsilon+m_{1s,3}\varepsilon^3+O(\varepsilon^5),
\end{equation}
with coefficient of $\varepsilon$ given by
\begin{equation}
\label{5.12}m_{1s,1}=\frac{\pi}{3a^2}\alpha\beta\big[(\alpha^6+\beta^6)\mu_1
+2\alpha^3\beta^3\mu_2\big],
\end{equation}
with factors
\begin{eqnarray}
\label{5.13}
\mu_1&=&\sum^6_{j=1}\big[w_j(d-(-1)^jw_j^2\alpha\beta)^2\big]^{-1}=
-36\frac{d^5\alpha^5\beta^5}{(d^6-a^3b^3)^2},\nonumber\\
\mu_2&=&\sum^6_{j=1}\big(-1)^j[w_j(d-(-1)^jw_j^2\alpha\beta)^2\big]^{-1}
=18d^2ab\frac{d^6+a^3b^3}{(d^6-a^3b^3)^2}.
\end{eqnarray}
The coefficient of $\varepsilon^3$ in Eq. (5.11) is given by
\begin{equation}
\label{5.14}
m_{1s,3}=\frac{\pi}{3}\alpha\beta\big[(\alpha^6+\beta^6)\mu_3+2\alpha^3\beta^3\mu_4\big],
\end{equation}
with factors
\begin{eqnarray}
\label{5.15}
\mu_3&=&\frac{3}{4}\sum^6_{j=1}\big[-w_j(d-(-1)^jw_j^2\alpha\beta)^4\big]^{-1}\nonumber\\
&=&-18d^3\alpha\beta a^2b^2\frac{14d^{12}+35d^6a^3b^3+5a^6b^6}{(d^6-a^3b^3)^4},\nonumber\\
\mu_4&=&\frac{3}{4}\sum^6_{j=1}(-1)^j\big[w_j(d-(-1)^jw_j^2\alpha\beta)^4\big]^{-1}\nonumber\\
&=&\frac{9}{2}ab\big[10d^{18}+125d^{12}a^3b^3+80d^6a^6b^6+a^9b^9\big]\big/(d^6-a^3b^3)^4.
\end{eqnarray}

Finally we consider the impetus. Expansion of the function $j_{1sm}$ in Eq. (4.2)
in powers of $\varepsilon$ yields
\begin{equation}
\label{5.16}j_{1sm}=j_{1sm,1}\varepsilon+j_{1sm,3}\varepsilon^3+O(\varepsilon^5),
\end{equation}
with coefficients
\begin{eqnarray}
\label{5.17}j_{1sm,1}&=&\pi\frac{b^3(a^3-b^3)}{a^3b^3-d^6},\nonumber\\
j_{1sm,3}&=&\frac{3\pi}{4}\frac{a^2b^3d^4(a^3-b^3)(5a^3b^3+7d^6)}{(a^3b^3-d^6)^3}.
\end{eqnarray}

Similarly we find for the function $\mathcal{I}_{1c}$ in Eq. (2.18)
\begin{equation}
\label{5.18}j_{1c}=j_{1c,1}\varepsilon+j_{1c,3}\varepsilon^3+O(\varepsilon^5),
\end{equation}
with the expressions
\begin{eqnarray}
\label{5.19}j_{1c,1}&=&12\pi\frac{d^2(a-b)}{a(4d^2-9ab)},\nonumber\\
j_{1c,3}&=&81\pi\frac{a^2b(a-b)(3ab+4d^2)}{(4d^2-9ab)^3}.
\end{eqnarray}

Substituting the above expansions into Eq. (4.3) we can compare with the complete
result and with the first
term in Eq. (5.1). As an example we consider the case $\eta=0.5,\;\rho=1,\;a=1,\;b=0.5,\;
d=3.5,\;\rho_a=\rho_b=1,\;\omega=1$.
The comparison in Fig. 2 shows that the perturbation expansion yields valid results
only up to $\varepsilon\approx 0.5$. For small $\varepsilon$ one finds in this case
$V^{(1)}=0.00201\;\varepsilon^2+O(\varepsilon^4)$. The velocity function $V^{(1)}$ increases
monotonically with $\varepsilon$.

\section{\label{VI}Small amplitude}
In this section we study the velocity function to second order on the amplitude $\varepsilon=f/a$.
First we formulate the asymptotic results for small amplitude $f$ and large distance $d$.
We find by substitution of the above results into Eq. (4.3) for the velocity function to order $f^2/d^2$
\begin{equation}
\label{6.1}V_{as}=\frac{27\xi^2}{1+\xi}\;\frac{\sigma_a^*-\xi^2\sigma_b^*}
{81(1+\xi)^2+4(\sigma_a^*+\xi^3\sigma_b^*)^2R^2}\;\frac{f^2}{d^2}R,
\end{equation}
with dimensionless variables
\begin{equation}
\label{6.2}\xi=b/a,\qquad\sigma_a^*=\sigma_a+\frac{1}{2},\qquad \sigma_b^*=\sigma_b+\frac{1}{2}.
\end{equation}
The result Eq. (6.1) holds for all $R=a^2\omega\rho/\eta$, small $f$ and large $d$.

For large viscosity $\eta$ or low frequency $\omega$ the behavior becomes
\begin{equation}
\label{6.3}V_{as}\approx\frac{1}{3}\;\frac{ab^2}{(a+b)^3}\;(a^2\rho_a^*-b^2\rho_b^*)\;
\frac{\omega f^2}{\eta d^2}\qquad (\mathrm{large}\; \eta, \mathrm{small}\; f\;\mathrm{and}\;\mathrm{large}\;d).
\end{equation}
For small viscosity or high frequency
\begin{equation}
\label{6.4}V_{as}\approx\frac{27}{4}\;\frac{ab^2}{a+b}\;
\frac{a^2\rho_a^*-b^2\rho_b^*}{(a^3\rho_a^*+b^3\rho_b^*)^2}
\;\frac{\eta f^2}{\omega d^2}\qquad (\mathrm{small}\;\eta, \mathrm{small}\; f \;\mathrm{and}\;\mathrm{large}\;d).
\end{equation}

The expressions in Eqs. (6.1-4) become identical to those derived earlier \cite{4},\cite{5} when $(\rho_a^*,\rho_b^*)$ is replaced by $(\rho_a,\rho_b)$, i.e. if the single sphere added mass is ignored. The expressions similar to Eqs. (6.3) and (6.4) derived by Derr et al.\cite{9} are the same apart from numerical factors. The potential flow appears only in the single sphere added mass. The dipolar interaction does not contribute in Eq. (6.1).  We refer to the Oseen model with single sphere added mass included as the Oseen* model. The velocity function $V^{(1)}_{O*}$ of this model is found from Eq. (4.3) with $j_{1sm},\;m_{0i},\;m_{1s}$ put equal to zero. The asymptotic behavior of the Oseen-Dipole model is the same as that of the Oseen* model.

The analytic form of Eq. (6.1) is the same as that encountered earlier in Eq. (4.5). Hence we conclude that for small amplitude $\varepsilon=f/a$ and long distance $d$ the velocity function is maximal at frequency given by
\begin{equation}
\label{6.5}R_{asx}=\frac{9}{2}\;\frac{1+\xi}{\sigma_a^*+\xi^3\sigma_b^*}.
\end{equation}
The corresponding value of the velocity function is
 \begin{equation}
\label{6.6}V_{asx}=\frac{3\xi^2}{4(1+\xi)^2}\;
\frac{\sigma_a^*-\xi^2\sigma_b^*}{\sigma_a^*+\xi^3\sigma_b^*}\;\frac{f^2}{d^2}.
\end{equation}
With these values the asymptotic results of Eq. (4.9) correspond to Eqs. (6.5) and (6.6).

One can choose $\xi$ such that $V_{asx}$ is maximized. This leads to a value $\xi_{asx}$ in the interval $0<\xi<1$ which is independent of $f/d$. In particular in case $\rho_a=\rho_b$ this becomes a root of the quartic equation $\xi^4-3\xi^3+2\xi^2-4\xi+2=0$ with the numerical value $\xi_{asx}=0.549$. In this case $\sigma_a^*=\sigma_b^*$ drops out in Eq. (6.6), so that the value of the velocity function maximum $V_{asx}^{(1)}$ is independent of the mass density $\rho_a=\rho_b$. The value of the maximum at $\xi_{asx}=0.549$ is $V_{asxx}^{(1)}=0.05647\varepsilon^2/\delta^2$, also independent of the mass density. The corresponding $R$-number is $R_{asxx}^{(1)}=5.981/\sigma_a^*$. For neutrally buoyant spheres $\sigma_a^*=3/2$ and $R_{asxx}^{(1)}=3.987$.

More generally we consider small amplitude $\varepsilon$, but any distance $d$. It is clear from Eq. (4.3) that the numerator is of order $\varepsilon^2$, whereas the denominator remains positive in the small $\varepsilon$ limit. To second order in $\varepsilon$ Eq. (4.5) becomes
\begin{equation}
\label{6.7}V_2=v_2\varepsilon^2=\frac{A_2R}{B_0+C_0R^2},
\end{equation}
with $A_2$ given by
\begin{equation}
\label{6.8}A_2=\frac{1}{6}z_{1s,1}\big[2\pi z_{0,0}(\sigma_a^*-2\xi^3\sigma_b^*)-3z_{0,0}j_{1sm,1}-3m_{0,0}j_{1c,1}\big]\varepsilon^2,
\end{equation}
with coefficients as given in the preceding section, and with $B_0$ and $C_0$ given by
\begin{equation}
\label{6.9}B_0=z_{0,0}^3,\qquad C_0=m_{0,0}^2z_{0,0}.
\end{equation}
We have dropped the superscript $(1)$, because the same result is obtained when higher order harmonics are included.
Eq. (6.7) is again of the form Eq. (4.5), so that we find values $R_{2x}$ and $V_{2x}$ as in Eq. (4.6),
\begin{equation}
\label{6.10} R_{2x}=\frac{z_{0,0}}{m_{0,0}},\qquad V_{2x}=\frac{A_2}{2z_{0,0}^2m_{0,0}}.
\end{equation}
From Eq. (5.3)
\begin{equation}
\label{6.11}z_{0,0}=6\pi(1+\xi)-18\pi\xi\frac{4\delta-3-3\xi}{4\delta^2-9\xi},
\end{equation}
and from Eqs. (3.16) and (5.8)
\begin{eqnarray}
\label{6.12_}m_{0,0}&=&\frac{4\pi}{3}(\sigma^*_a+\xi^3\sigma_b^*),\qquad (Oseen^*)\nonumber\\
m_{0,0}&=&\frac{4\pi}{3}(\sigma^*_a+\xi^3\sigma_b^*)
-2\pi\xi^3\frac{2\delta^3-1-\xi^3}{\delta^6-\xi^3},\qquad (Oseen-Dip).
\end{eqnarray}
The numerator $A_2$ in Eq. (6.10) reads for the Oseen*-model
\begin{equation}
\label{6.13}A_2=\varepsilon^2X[(2\delta-3)\sigma_a^*-(2\delta-3\xi)\xi^2\sigma_b^*],\qquad (Oseen^*),
\end{equation}
with prefactor
\begin{equation}
\label{6.14}X=576\pi^3\delta\xi^2\frac{(2\delta-3)(2\delta-3\xi)}{(4\delta^2-9\xi)^3}.
\end{equation}
Similarly for the Oseen-Dipole model
\begin{equation}\label{6.15}A_2=\varepsilon^2\frac{X}{2(\delta^6-\xi^3)}[r_a\sigma_a^*+r_b\sigma_b^*+r_0],\qquad (Oseen-Dip),
\end{equation}
with coefficients
\begin{eqnarray}\label{6.16}r_a&=&2(2\delta-3)(\delta^6-\xi^3),\nonumber\\
r_b&=&2\xi^2(\xi^2-\delta^2)(2\delta-3\xi)(\delta^4+\delta^2\xi+\xi^2),\nonumber\\
r_0&=&3\xi^2(1-\xi)(2\delta^4-3\xi+2\delta\xi-3\xi^2+2\delta\xi^2-3\xi^3).
\end{eqnarray}

We seek to optimize by finding the maximum value of $V_{2x}$ for fixed $\delta$. This leads to the optimum $\xi_x(\delta)$, i.e. the optimal radius of the second sphere for given length $d=\delta a$, and the corresponding maximum $V_{2xx}(\delta)$. For both models the value $\xi_x(\delta)$ is found as zero on the interval $0<\xi<1$ of a polynomial in $\xi$ of degree eight. In Fig. 3 we plot the function $\xi_x(\delta)$ for both models for the case of neutrally buoyant spheres.

In Fig. 4 we plot the corresponding functions $V_{2xx}(\delta)/\varepsilon^2$. The plots nearly coincide with the function $V_{asxx}/\varepsilon^2=0.05647/\delta^2$ found from the asymptotic calculation below Eq. (6.6).\\

As we noted below Eq. (6.9), the first harmonics approximation yields the exact small amplitude velocity function $V_2$ for the model considered. The higher order correction terms in Sec. V can be used to calculate the first harmonics approximation to the next order term $V_4=v_4\varepsilon^4$. We would need to include the second harmonics in order to find the exact $V_4$ for the model considered.

\section{\label{VII}Large amplitude}

The first harmonics approximation can be used also to perform model calculations for large amplitude of stroke. In this section we examine some situations of interest.

As a first example we show in Fig. 5 the velocity function $V$ of the
Oseen-Dipole model for the case
$\omega=1,\;a=1,\;b=0.5,\;d=3.5,\;\rho=\rho_a=\rho_b=1,\;f=2$ as a
function of the scaling number $R=a^2\omega\rho/\eta$, as
calculated from the exact expression Eq. (3.10) of Ref. 2, or
alternatively, from the expansion in harmonics with a sufficient
number of higher order harmonics (solid curve). The latter procedure is faster.
We have included up to order six.

We compare first with the asymptotic result Eq. (6.1) (short dashes),
as well as with this result with $\sigma_a^*$ replaced by $\sigma_a$ and$\sigma_b^*$ replaced by $\sigma_b$ (dot-dashed curve). The latter result is identical with that derived by Hubert et al.\cite{4}. For these values of $f$ and $d$ the asymptotic result is rather different from the exact one.

We compare with the result in first harmonics approximation, found from
Eq. (4.3) (long dashes). On the scale of the figure the latter curve also describes the result
for the Oseen$^*$ model, suggesting that for this value of $d$ the effect of the dipolar
interaction is small. The long-dashed curve is close to the solid one, showing that the first harmonics approximation is nearly exact.

The velocity function $V^{(1)}$ as a function of
$R_a=a^2\omega\rho_a/\eta$ and $S_a=\rho/\rho_a$ is also found
from Eq. (4.3). The expression is of the form Eq. (4.5) with $R$
replaced by $R_a$, with $A$ linear in $S_a$, $B$ independent of
$S_a$, and $C$ quadratic in $S_a$.

Next we consider a swimmer corresponding to the experiments and lattice Boltzmann simulations of Hubert et al. \cite{4} with parameters $a=8,\;b=5,\;d=28,\;\eta=1/6,\;
\rho=1,\;\rho_a=\rho_b=8$, and determine the velocity function maximum $V^{(1)}_x$ and corresponding number $R^{(1)}_x$ as functions of the stroke amplitude $\varepsilon=f/a$ for the Oseen-Dipole and the Oseen* model. From the asymptotic expressions Eqs. (6.5) and (6.6) we find for both models $R^{(1)}_{asx}=0.691$ and $V^{(1)}_{asx}=0.00444\;\varepsilon^2$. In Figs. 6  and 7 we plot the functions  $R^{(1)}_x(\varepsilon)$ and $V^{(1)}_x(\varepsilon)$ for the two models. The models yield nearly identical results, but there is a significant difference with the asymptotic values. In the limit $\varepsilon\rightarrow 0$ one finds from Eq. (6.10) for the Oseen-Dipole model $R_{2x}=0.5245$ and $V_{2x}=0.002194\;\varepsilon^2$, and for the Oseen* model  $R_{2x}=0.5236$ and $V_{2x}=0.002186\;\varepsilon^2$.

Returning to the general situation we seek to optimize the swimming speed by appropriate choice of
the ratio $\xi=b/a$ and the frequency $\omega$. For a start we consider
three cases with $d=3.5\;a$ fixed, but with different mass densities of the spheres. In the
first case $\rho_a=\rho_b=\rho=1$. In the second case
$\rho_a=\rho$, but $\rho_b=0$. A consideration of the contours of
$V^{(1)}$ in the $b\rho_a$-plane suggests that the latter choice
optimizes the velocity for fixed values of the other parameters.
In the third case both $\rho_a=0$ and $\rho_b=0$. We choose
amplitude $f=d-a-b$, the maximum amplitude for which there is no
overlap.

We use the notation $\sigma=\rho_b/\rho_a$. In the first case with $b=0.5\;a$ and $\sigma=1$
the velocity function takes the value $V_x^{(1)}=0.00885$ at the maximum at $R_{x}^{(1)}=3.117$. The exact value is $V_x=0.00859$.
In the second case with $b=0.5\;a$ and $\sigma=0$ we find
the value $V_x^{(1)}=0.0133$ at the maximum at $R_{x}^{(1)}=3.371$. The exact value is $V_x=0.0131$.
In the third case with $b=a$ and $\sigma=0$ we find
the value $V_x^{(1)}=0.0141$ at the maximum at $R_{x}^{(1)}=3.270$. The exact value is $V_x=0.0144$.

Next we consider a range of $d$ values. For each value of $\delta=d/a$ and given mass densities we choose the ratio $\xi=b/a$ and the
number $R$ such that $V^{(1)}$ is maximized. For these values
$\xi_x(\delta)$ we then calculate $V^{(1)}$ in the form Eq. (4.5) and
determine the values $R_{xx}^{(1)}$ and $V_{xx}^{(1)}$ as functions
of $\delta$. In the figures below we compare these functions for the Oseen-Dipole model with those for the Oseen$^*$ model.

We show first that for the Oseen$^*$ model with $\rho_a=\rho_b$ the maximum value $V^{[1]}_x$ has the simple property
that it is independent of the mass density $\rho_a=\rho_b$ of the spheres, as we found below Eq. (6.6) for the asymptotic situation. We note from Eq. (4.6) that $V^{[1]}_x=A/(2\sqrt{BC})$
with $A,B,C$ read off from Eq. (4.3). In the Oseen* model the moments $j_{1sm}$, $m_{0i}$, and $m_{1s}$ vanish. Hence in this model $B=z_0^3$,  $C=m_0^2z_0$ and
\begin{equation}
\label{7.1}V_x^{(1)}=\frac{A}{2m_0z_0^2},\qquad (Oseen^*)
\end{equation}
With the symmetry $\rho_a=\rho_b$ one has
\begin{eqnarray}
\label{7.2}A=\frac{2\pi}{3}z_{1s}\sigma^*_a[(1-\xi^3)\varepsilon z_0-2(1+\xi^3)j_{1cf}],\nonumber\\
m_0=\frac{4\pi}{3}\sigma_a^*(1+\xi^3),\qquad (Oseen^*,\rho_a=\rho_b).
\end{eqnarray}
In Eq. (7.1) the factor $\sigma_a^*$ cancels and the other coefficients do not depend on the mass densities, so that the velocity function maximum $V_x^{(1)}$ is independent of the mass density $\rho_a=\rho_b$ of the spheres. It follows that then also $\xi_x(\delta)$ is independent of $\rho_a=\rho_b$.
The above properties hold also for the Oseen model.

We choose amplitude $f=d-2a$. In Fig. 8 we show the function $\xi_x(\delta)$ for the case
$\rho_a=\rho_b=\rho$ for both the Oseen-Dipole model and the
Oseen* model. In Fig. 9 we show the corresponding functions
$R_{xx}^{(1)}(\delta)$ and in Fig. 10 we show the functions $V_{xx}^{(1)}(\delta)$.
In particular, at $\delta=3.5$ we have $\xi_x(3.5)=0.570$, $R_{xx}^{(1)}(3.5)=3.054$ , $V_{xx}^{(1)}(3.5)=0.00533$ for the Oseen-Dipole model and $\xi_x(3.5)=0.561$, $R_{xx}^{(1)}(3.5)=3.032$ , $V_{xx}^{(1)}(3.5)=0.00516$ for the Oseen* model.

According to the above argument for other values of $\rho_a=\rho_b$ the curves for $\xi_x(\delta)$ and  $V_{xx}^{(1)}(\delta)$ for the Oseen* model
are the same as shown in Figs. 9 and 10. The corresponding curves for the Oseen-Dipole model are not much different. In Figs. 11 and 12
 we show the curves
for $\xi_x(\delta)$ and $R_{xx}^{(1)}(\delta)$ at $\rho_a=\rho_b=0.5\rho$ for both models. The curve for $V_{xx}^{(1)}(\delta)$  for the Oseen* model is the same as the one in Fig. 10, and the one for the Oseen-Dipole model is quite similar.
In particular at $\delta=3.5$ we have $\xi_x(3.5)=0.574,\;R_{xx}^{(1)}(3.5)=4.555,\; V_{xx}^{(1)}(3.5)=0.00542$ for the Oseen-Dipole model and $\xi_x(3.5)=0.561,\;R_{xx}^{(1)}(3.5)=4.529,\; V_{xx}^{(1)}(3.5)=0.00516$ for the Oseen* model.

Consider again the Oseen-Dipole swimmer corresponding to Fig. 5, $a=1,\;b=0.5,\;d=3.5$, but with amplitude of stroke $f=1.5$. We choose units such that the fluid has viscosity $\eta=1$ and mass density $\rho=1$. If the mass density of both spheres is $\rho_a=\rho_b=1$, then the two-sphere in first harmonics approximation is found from Eq. (4.6) to have velocity function maximum $V^{(1)}_{x}=0.00513$, achieved at $R$-number $R_{x}=3.129$. As we showed above,
if we increase the radius of the second sphere to $\xi_{x1}=0.570$ then the velocity function maximum increases to $V^{(1)}_{xx}=0.00533$, achieved at $R$-number $R_{xx}=3.054$. If instead we first decrease the mass density of the spheres to $\rho_a=\rho_b=0.5$, and increase the radius of the second sphere to $\xi_{x2}=0.574$, then we achieve velocity function maximum $V^{(1)}_{xx}=0.00542$, at $R$-number $R_{xx}=4.555$.

\section{\label{VIII}Power and efficiency}

In the preceding sections we concentrated on the calculation of the mean swimming velocity.
Other properties of both the Oseen* model and the Oseen-Dipole model can also be investigated in full detail. In this section
we consider the power dissipated per period and the efficiency of swimming. The latter is
defined as the ratio of speed and power.

In both models the relative distance between both centers is prescribed as
$x_2(t)-x_1(t)=d+f\sin(\omega t)$. The center position
$C(t)=\frac{1}{2}(x_1(t)+x_2(t))$ is found by integration of the swimming
velocity $U_{sw}(t)=dC(t)/dt$. Hence the two sphere centers move as
$x_1(t)=C(t)-(d+f\sin(\omega t))/2$ and $x_2(t)=C(t)+(d+f\sin(\omega t))/2$
The two sphere velocities are $U_1(t)=U_{sw}(t)-\frac{1}{2}\omega f \cos(\omega t)$ and
$U_2(t)=U_{sw}(t)+\frac{1}{2}\omega f \cos(\omega t)$.

The time-dependent rate of dissipation is given by
\begin{equation}
\label{8.1}\mathcal{D}=\du{U}\cdot\vc{\zeta}\cdot\du{U},
\end{equation}
where $\du{U}=(U_1,U_2)$. The power used for a stroke in periodic swimming is the
average dissipation in a period,
\begin{equation}
\label{8.2}P=\overline{\mathcal{D}}.
\end{equation}
The efficiency is defined as \cite{12}
\begin{equation}
\label{8.3}L=8\pi\eta\omega a^2\frac{\overline{U_{sw}}}{\overline{\mathcal{D}}}.
\end{equation}
The dimensionless mean rate of dissipation $\hat{\mathcal{D}}$ is defined by
\begin{equation}
\label{8.4}\overline{\mathcal{D}}=8\pi\eta\omega^2 a^3\hat{\mathcal{D}},
\end{equation}
so that $L=V/\hat{\mathcal{D}}$.

The mean rate of dissipation $\hat{\mathcal{D}}$ can be calculated in first harmonics
approximation in analytic form, but the expression for $\hat{\mathcal{D}}^{(1)}$ becomes too
complicated to be practically useful. The required values can be calculated from the numerical
calculation as before \cite{11}.

Depending on the starting point and the size of the shift the change of
$V^{(1)}$ upon doubling the viscosity or the mass density can be positive, negative, or zero.
In Table I we list values of $V^{(1)}$ and $L^{(1)}$ for selected values of $R_a=a^2\omega\rho_a/\eta$ and $S_a=\rho/\rho_a$. These
can be compared with the corresponding values of $V$ and $L$ in Table I of Ref. 11, calculated
by inclusion of higher order harmonics. In the latter Table some copying errors need to be
corrected. The last but third column of that Table should have heading $10^5L$ and entries
$631,1000,873$.  Comparison of the two Tables shows that for this large amplitude the approximate value $V^{(1)}$ differs from the exact $V$ by about ten percent. Table I shows that the efficiency varies significantly over the $R_aS_a$-plane. We note that $R=R_aS_a$.

\begin{table}[!htb]  \footnotesize\centering
  \caption{}\label{tab:1}
\begin{tabular}{|c|c|c|c|c|c|c|c|c|c|c|c|}
\hline
 $R_a,S_a$ &$10^6V^{(1)}$ &$10^6L^{(1)}$& $R_a,S_a$ &$10^5V^{(1)}$ &$10^5L^{(1)}$&$R_a,S_a$
 &$10^5V^{(1)}$ &$10^5L^{(1)}$& $R_a,S_a$ &$10^6V^{(1)}$ &$10^5L^{(1)}$
 \rule[-5pt]{0pt}{16pt} \\
\hline
 $0.1,1$ &$ 569$ &$ 752$ & $1,1$ &$ 516$ &$ 679$ &$10,1$ &$ 504$ &$ 638$ & $100,1$ &$ 553$ &$70$

 \rule[-5pt]{0pt}{16pt} \\
 $0.05,1$ &$ 284$ &$ 376$ & $0.5,1$ &$ 277$ &$ 367$ &$5,1$ &$ 797$ &$ 1017$ & $50,1$ &$ 1102$ &
 $ 139$
 \rule[-5pt]{0pt}{16pt} \\
 $0.05,2$ &$ 386$ &$ 508$ & $0.5,2$ &$ 367$ &$ 486$ &$5,2$ &   $ 702$ &$886$ & $50,2$ &$ 856$ &
 $ 107$
\rule[-5pt]{0pt}{16pt} \\

 \hline
\end{tabular}
\end{table}-

\subsection*{}
List of values of the velocity function $V^{(1)}(R_a,S_a)$ and the efficiency function
$L^{(1)}(R_a,S_a)$ for the Oseen-Dipole model of the two-sphere with $\xi=0.5,\;\delta=3.5,\;\varepsilon=2,\;\rho_a=\rho_b$.\\\\

In Ref. 11 we found for a particular swimmer that the surface $L(R_a,S_a)$ looks
quite similar to the surface $V(R_a,S_a)$.
This implies that $\hat{\mathcal{D}}$ is nearly constant in the $R_aS_a$-plane.
In Fig. 13 we show the
surface $\hat{\mathcal{D}}^{(1)}(R_a,S_a)$ for the case $\xi=0.5,\;\delta=3.5,\;\varepsilon=2,\;\rho_a=\rho_b=\rho$.

\section{\label{IX}Discussion}

The study shows  that the first harmonics approximation provides a useful tool for the analysis of
the two-sphere Oseen-Dipole model and of the Oseen* model. The approximation yields an analytic expression for the mean swimming velocity which can be studied in its dependence on parameters. The swimming speed can be compared with the exact one as found from another expression derived earlier \cite{2}, or as found numerically from an expansion including higher order harmonics. The comparison in special cases with large amplitude of stroke shows that the first harmonics approximation leads to results in good agreement with the exact ones.
For small amplitude the first harmonics approximation becomes exact. In Sec. VI we derived the explicit expression for the small amplitude mean swimming velocity for both models.

The first harmonics expression for the mean swimming velocity Eq. (4.3) shows a simple dependence on the scaling numbers $R=a^2\omega\rho/\eta$ and $S_a=\rho/\rho_a$, consisting of a Pad\'e type ratio of two simple polynomials in terms of the two numbers. The dependence on $R$ shows that the mean swimming velocity in any chosen fluid varies slowly over a wide range of frequency of the stroke. The expression allows us to optimize the swimming velocity by adjusting the frequency and the ratio of radii.

The Oseen-Dipole model is more sophisticated than the Oseen$^*$ model since it includes the dipolar interactions of potential flow. However, it fails to describe the interesting reversal of swimming direction at high frequency seen by Dombrowski and Klotsa \cite{8} in a numerical analysis based on the Navier-Stokes equations.
We view the present study as a necessary prologue to further investigation. We expect that the model can be adapted with more realistic hydrodynamic interactions to describe the reversal phenomenon.\\\\

The author has no conflicts of interest to disclose.

\newpage
\clearpage
\newpage
\setlength{\unitlength}{1cm}
\begin{figure}
 \includegraphics{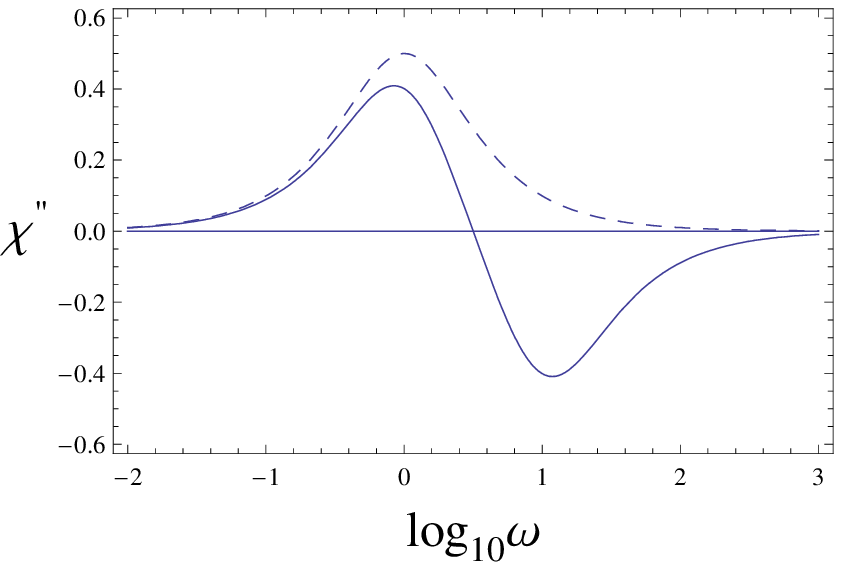}
   \put(-9.1,3.1){}
\put(-1.2,-.2){}
  \caption{Plot of $\chi''(\omega)$ given by Eq. (4.12) (dashed curve) and of $\chi''_2(\omega)$ given by Eq. (4.13) (solid curve) for the numerical example $\chi_0=1,\;\gamma=1,\;\chi_{02}=-1,\;\gamma_2=10$.
}
\end{figure}
\newpage
\setlength{\unitlength}{1cm}
\begin{figure}
 \includegraphics{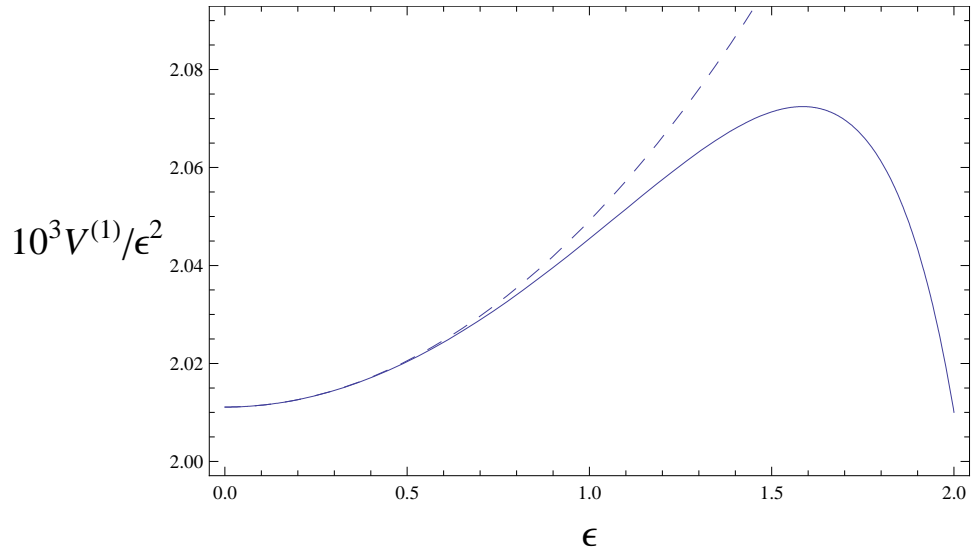}
   \put(-9.1,3.1){}
\put(-1.2,-.2){}
  \caption{Plot of the velocity function $V^{(1)}$ as a function of amplitude $\varepsilon=f/a$ for the case $\eta=0.5,\;\rho=1,\;a=1,\;b=0.5,\;
d=3.5,\;\rho_a=\rho_b=1,\;\omega=1$ (solid curve) and of the approximation given by the expansion of
coefficients in powers of $\varepsilon$ as derived in Sec. V (dashed curve). On the scale of the figure the dashed curve cannot be distinguished from the expansion up to the quartic term, as given by Eq. (5.1).}
\end{figure}

\newpage
 \setlength{\unitlength}{1cm}
\begin{figure}
 \includegraphics{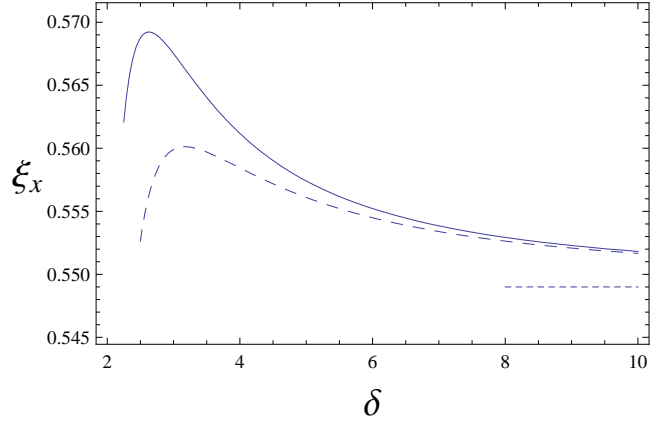}
   \put(-9.1,3.1){}
\put(-1.2,-.2){}
  \caption{Plot of the function $\xi_x(\delta)$ in the small amplitude limit for a two-sphere with mass densities $\rho_a=\rho_b=\rho$ for the Oseen-Dipole model (solid curve) and the Oseen*-model (long dashes). The horizontal line (short dashes) is the asymptotic limit $\xi_{asx}=0.549$.}
\end{figure}

\newpage
 \setlength{\unitlength}{1cm}
\begin{figure}
 \includegraphics{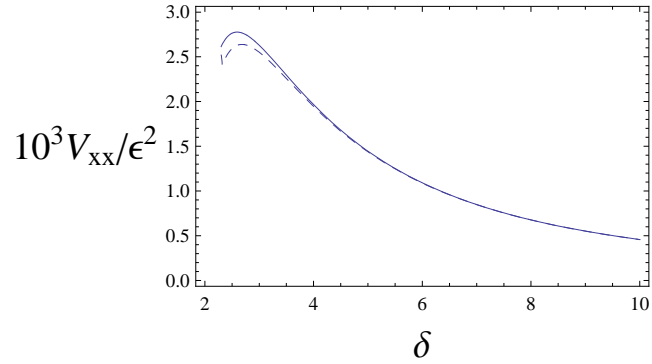}
   \put(-9.1,3.1){}
\put(-1.2,-.2){}
  \caption{Plot of the optimal velocity function $V_{2xx}(\delta)$ for a two-sphere with mass densities $\rho_a=\rho_b=\rho$ for the Oseen-Dipole model (solid curve) and the Oseen*-model (long dashes). The plots nearly coincide with the function $V_{asxx}/\varepsilon^2=0.05647/\delta^2$ found from the asymptotic calculation below Eq. (6.6).}
\end{figure}

\newpage
\setlength{\unitlength}{1cm}
\begin{figure}
 \includegraphics{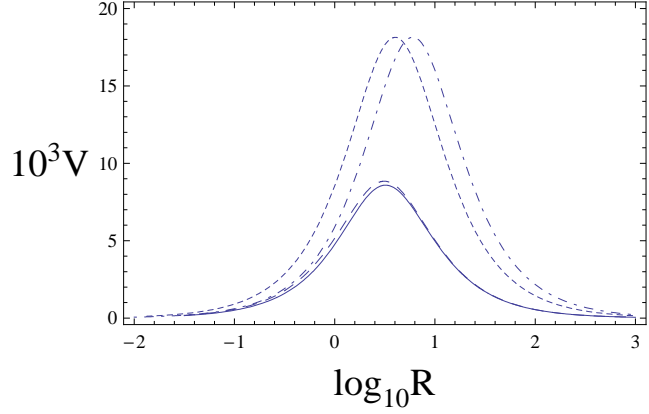}
   \put(-9.1,3.1){}
\put(-1.2,-.2){}
  \caption{Plot of the velocity function $V$ for parameters $\xi=b/a=0.5,\;\delta=d/a=3.5,\;\varepsilon=f/a=2,\;\rho_a=\rho_b=\rho=1$ as a function of $R=a^2 \omega\rho/\eta$ (solid curve). We compare with the velocity function $V^{(1)}$ calculated in first harmonics approximation (short dashes) and with the asymptotic result Eq. (6.1) (long dashes), and the analogous result for the Oseen model \cite{4} (dot-dashed curve).}
\end{figure}
\newpage
\setlength{\unitlength}{1cm}
\begin{figure}
 \includegraphics{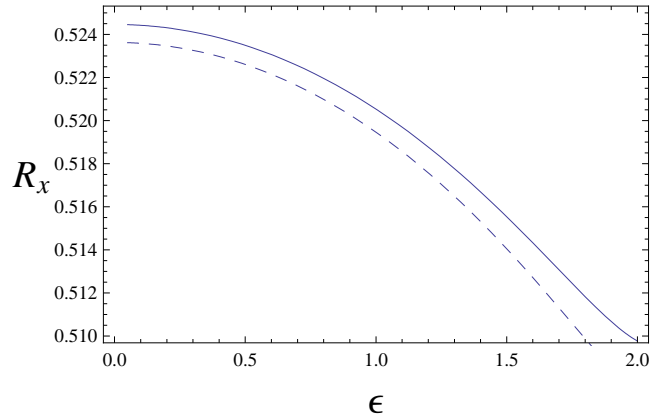}
   \put(-9.1,3.1){}
\put(-1.2,-.2){}
  \caption{Plot of the function $R^{(1)}_x(\varepsilon)$  for a two-sphere with parameters $a=8,\;b=5,\;d=28,\;\eta=1/6,\;
\rho=1,\;\rho_a=\rho_b=8$, as a function of $\varepsilon=f/a$ for the Oseen-Dipole model (solid curve) and the Oseen$^*$ model (dashed curve).}
\end{figure}

\newpage
\setlength{\unitlength}{1cm}
\begin{figure}
 \includegraphics{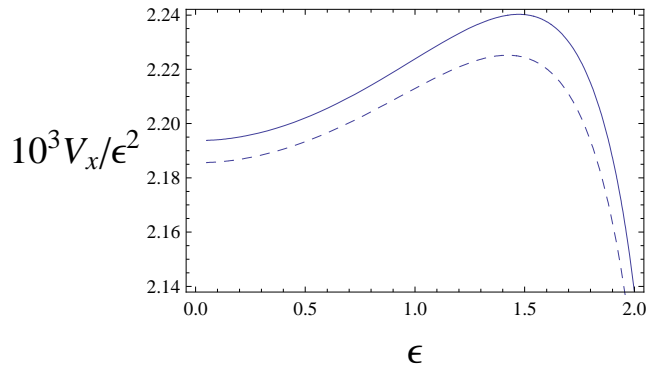}
   \put(-9.1,3.1){}
\put(-1.2,-.2){}
  \caption{As in Fig. 6 for the function $V^{(1)}_x(\varepsilon)$.
}
\end{figure}

\newpage
\setlength{\unitlength}{1cm}
\begin{figure}
 \includegraphics{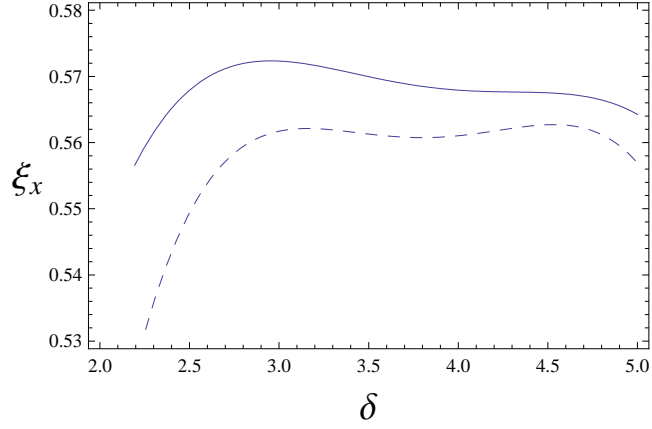}
   \put(-9.1,3.1){}
\put(-1.2,-.2){}
  \caption{Plot of the function $\xi_x(\delta)$ representing the ratio $b/a$ with optimal  $V^{(1)}_x$ for a two-sphere with mean distance $d$ between centers and
stroke of amplitude $\varepsilon=\delta-1-\xi_x$, as a function of $\delta=d/a$ for the Oseen-Dipole model (solid curve) and the Oseen$^*$ model (dashed curve) with mass densities $\rho_a=\rho_b=\rho$.}
\end{figure}
\newpage
\setlength{\unitlength}{1cm}
\begin{figure}
 \includegraphics{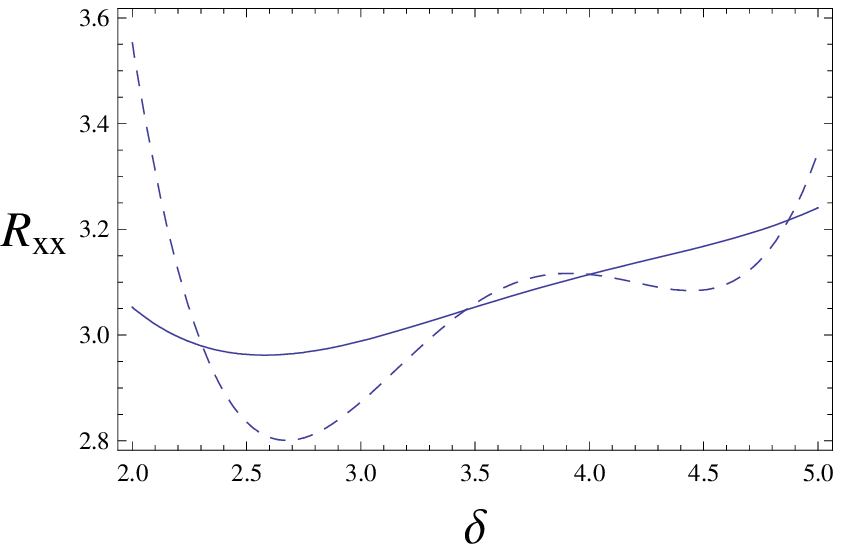}
   \put(-9.1,3.1){}
\put(-1.2,-.2){}
  \caption{Plot of the function $R^{(1)}_{xx}(\delta)$ corresponding to the optimum specified in caption 8.}
\end{figure}
\newpage
\setlength{\unitlength}{1cm}
\begin{figure}
 \includegraphics{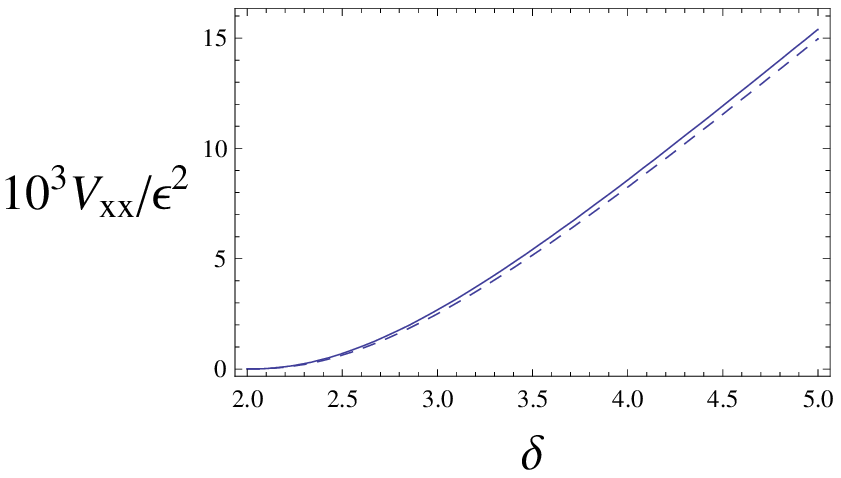}
   \put(-9.1,3.1){}
\put(-1.2,-.2){}
  \caption{Plot of the function $V^{(1)}_{xx}(\delta)$ corresponding to the optimum  specified in caption 8.}
  \end{figure}

  \newpage
\setlength{\unitlength}{1cm}
\begin{figure}
 \includegraphics{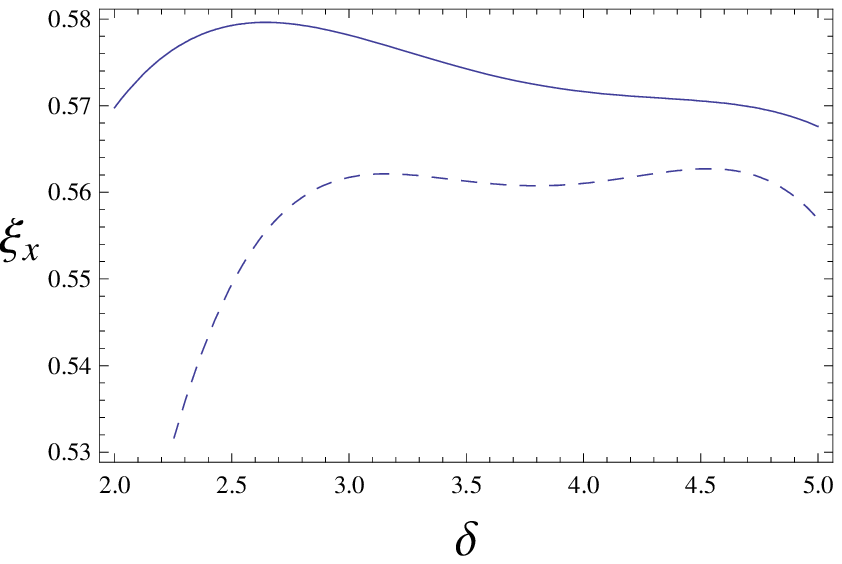}
   \put(-9.1,3.1){}
\put(-1.2,-.2){}
  \caption{As in Fig. 8 for a two-sphere with mass densities $\rho_a=\rho_b=0.5\rho$.}
\end{figure}

\newpage

\setlength{\unitlength}{1cm}
\begin{figure}
 \includegraphics{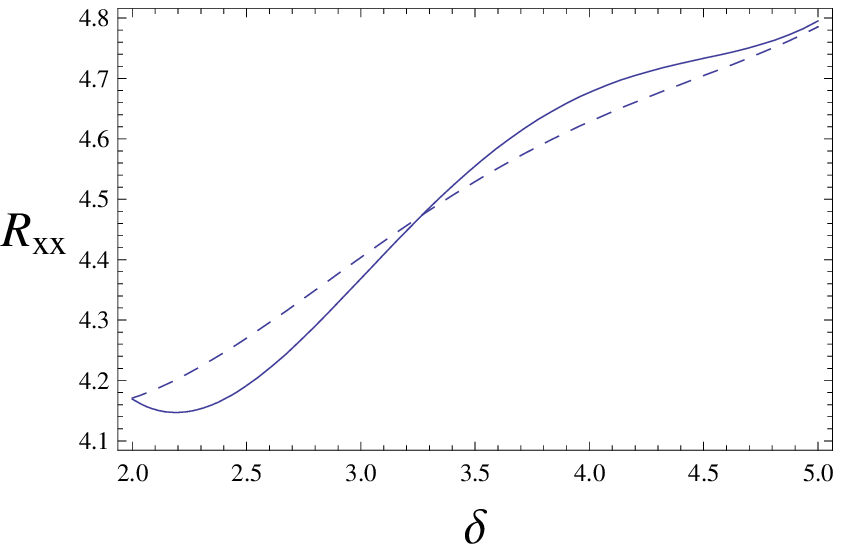}
   \put(-9.1,3.1){}
\put(-1.2,-.2){}
  \caption{As in Fig. 9 for a two-sphere with mass densities $\rho_a=\rho_b=0.5\rho$.}
  \end{figure}

\newpage

\setlength{\unitlength}{1cm}
\begin{figure}
 \includegraphics{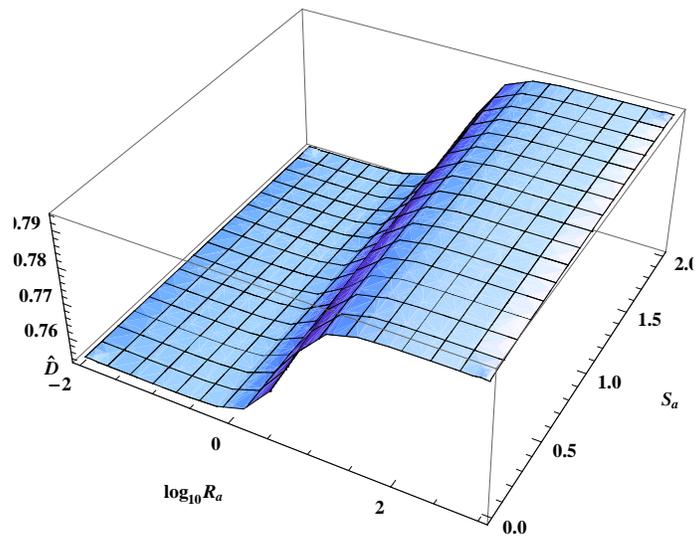}
   \put(-9.1,3.1){}
\put(-1.2,-.2){}
  \caption{Plot of the
surface $\hat{\mathcal{D}}^{(1)}(R_a,S_a)$, as defined by Eqs. (8.1-4), for the case $\xi=0.5,\;\delta=3.5,\;\varepsilon=2,\;\rho_a=\rho_b=\rho$.}
\end{figure}

\end{document}